\documentclass[aps,twocolumn,showpacs,amsfonts,amsmath,superscriptaddress,amssymb,prl]{revtex4}

\usepackage{graphicx}
\usepackage{dcolumn}
\usepackage{bm}
\usepackage{longtable}

\begin{document}

\title{Superconducting transition at $\rm 38~K$ in insulating-overdoped $\rm La_{2}CuO_{4}-La_{1.64}Sr_{0.36}CuO_{4}$ superlattices:
Evidence for interface electronic redistribution from resonant soft x-ray scattering}

\author{S. Smadici}
  \affiliation{Frederick Seitz Materials Research Laboratory, University of Illinois, Urbana, IL 61801, USA}%

\author{J. C. T. Lee}
  \affiliation{Frederick Seitz Materials Research Laboratory, University of Illinois, Urbana, IL 61801, USA}%

\author{S. Wang}
  \affiliation{Frederick Seitz Materials Research Laboratory, University of Illinois, Urbana, IL 61801, USA}%

\author{P. Abbamonte}
  \affiliation{Frederick Seitz Materials Research Laboratory, University of Illinois, Urbana, IL 61801, USA}%

\author{G. Logvenov}
  \affiliation{Brookhaven National Laboratory, Upton, NY 11973, USA}%

\author{A. Gozar}
  \affiliation{Brookhaven National Laboratory, Upton, NY 11973, USA}%

\author{C. Deville Cavellin}
  \affiliation{Brookhaven National Laboratory, Upton, NY 11973, USA}%
  \affiliation{Universit$\acute{e}$ Paris 12, 94010 Cr$\acute{e}$teil Cedex, France}%

\author{I. Bozovic}
  \affiliation{Brookhaven National Laboratory, Upton, NY 11973, USA}%

\begin{abstract}
We use resonant soft x-ray scattering (RSXS) to quantify the hole
distribution in a superlattice of insulating $\rm La_{2}CuO_{4}$
(LCO) and overdoped $\rm La_{2-x}Sr_{x}CuO_{4}$ (LSCO). Despite its
non-superconducting constituents, this structure is superconducting
with $\rm T_{c}=38~K$. We found that the conducting holes
redistribute electronically from LSCO to the LCO layers. The LCO
layers were found to be optimally doped, suggesting they are the
main drivers of superconductivity. Our results demonstrate the
utility of RSXS for separating electronic from structural effects at
oxide interfaces.

\end{abstract}


\maketitle

The interface between two correlated electron systems can exhibit
ground states that are not stable in the bulk of either of its
constituents. This tendency could provide a novel route to new
devices.~\cite{altieri,pickett,okamoto2004,okamoto2006,C2007} An
interesting realization of this idea are the two-dimensional
electron gases (2DEGs) observed at the interfaces between LaTiO$_3$
and SrTiO$_3$\cite{OMGH2002}, SrTiO$_3$ and
LaAlO$_3$\cite{ohtomo2004,RTCF2007}, LaMnO$_3$ and
SrMnO$_3$\cite{koida2002,S2007,BMTW2008}, La$_2$CuO$_4$ and
La$_{2-x}$Sr$_x$CuO$_4$\cite{G2008,BLVC2004}, and ZnO and
Mg$_x$Zn$_{1-x}$O\cite{Tsukazaki2007}, which exhibit phenomena
ranging from magnetoresistance to the quantum Hall effect to
superconductivity.

One of the outstanding questions in this field is whether the 2DEGs
observed in these heterostructures are truly `intrinsic' interface
properties that arise from charge accumulation, or if they simply
arise from defects like oxygen vacancies or cation interdiffusion.
These two effects are virtually impossible to distinguish with
transport measurements alone.

In this paper we present a quantitative study, using resonant soft
x-ray scattering (RSXS), of the charge distribution in a
superlattice of insulating $\rm La_{2}CuO_{4}$ (LCO) and overdoped
$\rm La_{1.64}Sr_{0.36}CuO_{4}$ (LSCO). Despite its
nonsuperconducting constituents, this heterostructure is
superconducting with $\rm T_{c} = 38~K$.~\cite{G2008} Using RSXS, a
technique that can probe the holes independently of the atomic
lattice, we find that the hole density varies more gradually than
the distribution of Sr$^{2+}$ cations, indicating redistribution of
carriers among the layers. Applying a linear response model, we show
that this redistribution takes place over a characteristic distance
of $\lambda_0=6.1\pm 2.0~\rm\AA$. The LCO layers are found to be
highly doped, with filling $p=0.18$ holes/Cu, suggesting that
superconductivity arises in the `insulating' substructure. Our
results show that genuine charge accumulation can be achieved in
oxide heterostructures, and can be observed with RSXS.

\begin{figure}
\centering\rotatebox{0}{\includegraphics[scale=0.4]{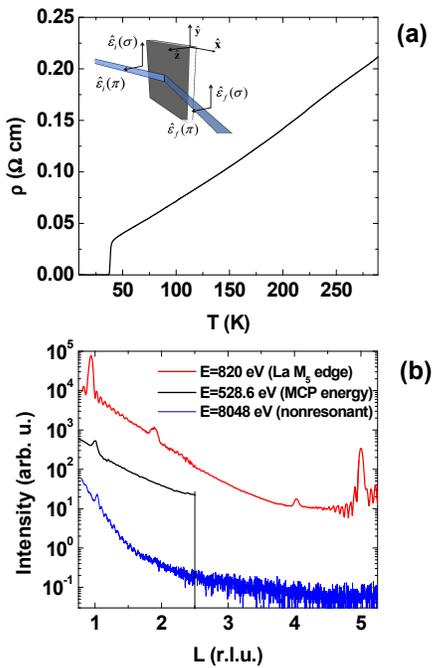}}
\caption{\label{fig:Figure1} (color online) (a). Superlattice
resistivity showing $\rm T_{c}=38.4~K$. (Inset) Experimental
geometry showing the direction of incident and scattered
polarizations. (b). Specular x-ray reflectivity measurements as a
function of $L$ (blue) off resonance, (red) near the $\rm La~M_{5}$
edge, and (black) at the mobile carrier peak (MCP) below the O $K$
edge. At the La $\rm M_{5}$ edge the peaks are slightly shifted from
the integer values because of refraction effects. Thickness
oscillations are visible, indicating flat interfaces. The width of
the $L=1$ peak is determined by the thickness of the superlattice.
The $L=2$ reflection is visible at the La edge, but not at the MCP,
indicating that the holes do not follow the profile of Sr dopants.}
\end{figure}

\begin{figure}
\centering\rotatebox{-0}{\includegraphics[scale=0.4]{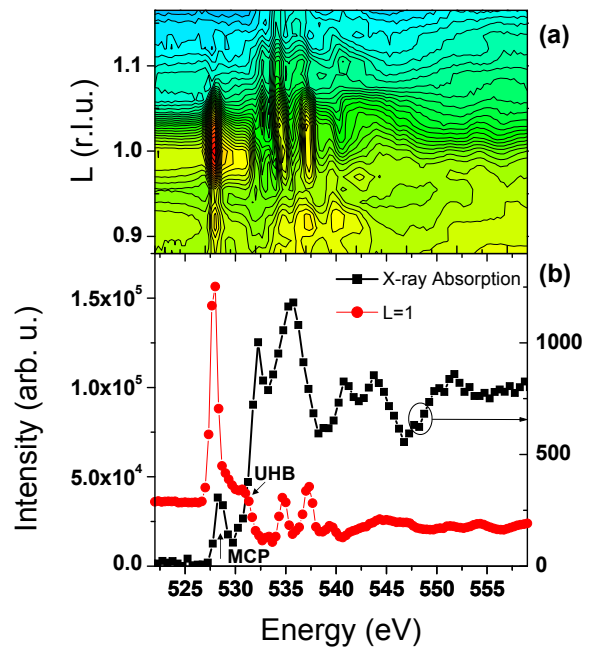}}
\caption{\label{fig:Figure2} (color online) Energy dependence of the
$L=1$ superlattice reflection near the O $K$ edge. (a). Color plot
(logarithmic color scale), showing that the reflection is enhanced
at the energy of the mobile carrier prepeak (MCP), which indicates
that the holes are modulated with the period of the superlattice.
(b). Summary plot comparing the intensity of the $L=1$ reflection
from (a) (red circles), to the x-ray absorption spectrum (black
squares).  The slight energy shift between the XAS and resonance
maximum occurs because of interference between resonant and
off-resonant scattering. This interference allows us to determine
quantitatively the amplitude of the hole modulation. }
\end{figure}

Superlattices with LCO and LSCO sublayers of various thicknesses
were grown in a unique atomic layer-by-layer MBE (ALL-MBE) system on
$\rm LaSrAlO_{4}$ (LSAO) substrates. Structures were monitored
during growth with time-of-flight ion scattering and recoil
spectroscopy (TOF-ISARS) and reflection high-energy electron
diffraction (RHEED). The samples were annealed to remove any excess,
interstitial oxygen. The single-phase overdoped LSCO has in-plane
lattice constants almost the same (to within $\rm 0.03~\%$) as the
LSAO substrate. LCO layers are strained, but even under maximal
strain single-phase LCO remains insulating and not superconducting.
The sample selected for RSXS studies consisted of 15 repeats of 2
$\times$ LCO + 4 $\times$ LSCO, which despite its nonsuperconducting
constituents had $\rm T_{c}=38~K$,~\cite{G2008} close to the value
for optimally doped LSCO crystals.  Hard x-ray measurements
(Fig.~\ref{fig:Figure1}b, blue line) were done on a reflectometer.
X-ray absorption (XAS) measurements (Fig.~\ref{fig:Figure2}b, square
symbols) showed the sample to be highly doped on average, with a
small peak at the upper Hubbard band (UHB) that is seen more clearly
in the resonance profile (red symbols).

RSXS measurements were done at the undulator beam line X1B at the
National Synchrotron Light Source in an ultra-high vacuum (UHV)
diffractometer. Measurements were made in the specular geometry,
i.e. in reflectivity mode, with the momentum transfer perpendicular
to the plane of the superlattice. Momenta will be written in terms
of the third Miller index $L$, i.e. $Q=2\pi \rm{L}/\emph{c}$, where
$c=39.84~\rm\AA$ is the superlattice period. The X1 undulator
produces both horizontally and vertically polarized light in the
proportions $P_{\pi}=0.93$ and $P_{\sigma}=0.07$, defined by the
orientation with respect to the scattering plane
(Fig.~\ref{fig:Figure1}a, inset). The detector integrated over both
scattering channels. XAS measurements were done {\it in situ} in
fluorescence yield mode. The incident energy resolution was set to
$\rm \Delta E = 0.2~eV$, and all measurements were done at $\rm T =
90~K$ which was found to eliminate radiation damage.

Initial RSXS measurements are summarized in Fig.~\ref{fig:Figure1}b,
which shows the scattered intensity as a function of $L$ for various
photon energies. The data are rather featureless far from resonance
($\rm 8048~eV$). Near the La $M_5$ edge ($\rm 820~eV$) however,
several peaks are visible at integer $L$, which are reflections from
the superlattice period \cite{note2}. The reason these are visible
at the La edge is that the contrast between LCO and LSCO layers,
which have different La content, is enhanced. The relative
intensities of these peaks are determined by the profile of Sr
dopants in the superlattice.

The scattering of x-rays from the doped holes, which reflects their
distribution, is enhanced if the beam energy is tuned to the mobile
carrier peak (MCP) below the O $K$ edge~\cite{AVRS2002,ARSG2005}
(Fig.~\ref{fig:Figure2}b, arrow). In Figs.~\ref{fig:Figure2}a and 2b
(red circles) we show scattering near this energy ($\rm 528.6~eV$).
A giant resonance is visible near $L=1$, indicating that, as
expected, the holes are modulated with the period of the
superlattice.~\cite{Footnote5} In contrast to scattering at the La
edge, however, no peak is visible at $L=2$ for scattering at the MCP
energy. (Fig.~\ref{fig:Figure1}b) This shows that the distribution
of holes does not exactly follow that of the Sr atoms. Suppression
of a harmonic, in fact, suggests that the hole density varies more
gradually than the profile of the Sr$^{2+}$ ions, which is evidence
for redistribution of holes among the layers.

The hole distribution can be determined quantitatively from this
scattering. The integrated intensity of the charge reflection at
momentum $L$ is given by
\begin{eqnarray}
\frac{I_{L}}{AV}=P_{\sigma}|\hat{\epsilon}_{f}^{*}(\sigma) \cdot
\textbf{S}_{L} \cdot \hat{\epsilon}_{i}(\sigma)|^2 + P_{\pi}
|\hat{\epsilon}_{f}^{*}(\pi) \cdot \textbf{S}_{L} \cdot
\hat{\epsilon}_{i}(\pi)|^2 \label{eqn:Equation1}
\end{eqnarray}
\noindent where $\rm {\textbf{S}}_\emph{{L}}$ is the structure
factor tensor, $\hat{\epsilon}_{f}$ and $\hat{\epsilon}_{i}$ are
final and initial polarization states (defined as in
Fig.~\ref{fig:Figure1}a, inset), $V$ is the scattering volume, and
$A$ contains all proportionality factors, e.g. beam intensity, unit
cell volume, etc. The $\sigma$ term must be included since, although
$P_{\sigma}$ is small, $\hat{\epsilon}_{i,f}(\sigma)$ lie in the
$\rm CuO_{2}$ plane so this scattering has a large resonant
enhancement.

The structure factor comprises two terms
$\textbf{S}_{L}=\textbf{S}_L^0+\textbf{S}^{D}_{L}$. The first term
\begin{eqnarray}
(\textbf{S}_L^0)_{ij}=\delta_{ij} \sum_{l,n}
\,d^{n}_{l}\,f_{n}(\omega)\,e^{i2\pi L z_{l}/c}
\label{eqn:Equation1a}
\end{eqnarray}
\noindent is the lattice structure factor, which is isotropic and
describes scattering from the atomic lattice. Here $z_{l}$ is the
position of layer $l$, $d^{n}_{l}$ is the number of atoms of type
$n$ in one $a\times a$ area of this layer, and $f_{n}(\omega)$ is
the scattering factor of atom type $n$.~\cite{Footnote4} The matrix
$d^{n}_{l}$ defines the structure of the superlattice, including the
distribution of defects. For clarity we will first analyze our data
assuming a perfect structure, and correct the result for defects
afterward.

The second term
\begin{eqnarray}
(\textbf{S}^{D}_{L})_{ij}=f^{D}_{ij}(\omega) \,\sum_{l}
\,\frac{p_{l}^{0}}{2}\,e^{i2\pi L z_{l}/c} \label{eqn:Equation1b}
\end{eqnarray}
\noindent is anisotropic and describes scattering from the doped
holes. $p_{l}^{0}$ is the hole count per Cu atom in layer $l$ in an
ideal structure, and $f^{D}_{ij}(\omega)$ is the scattering power of
a doped hole.~\cite{ARSG2005,A2006} The factor $1/2$ accounts for
the fact that there are two planar oxygen atoms for each Cu atom.

\begin{figure}
\centering\rotatebox{-90}{\includegraphics[scale=0.4]{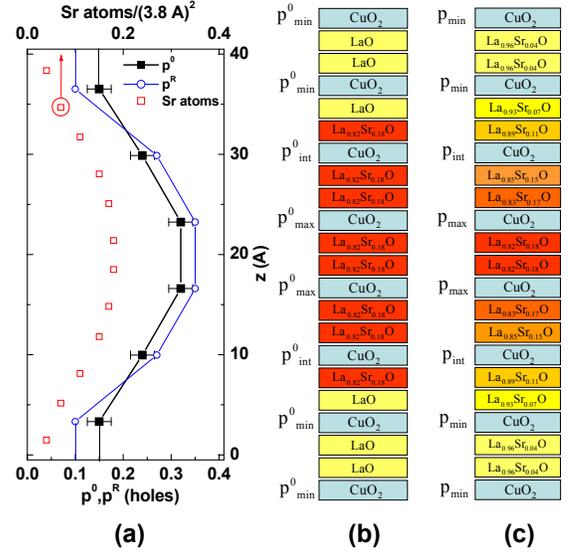}}
\caption{\label{fig:Figure3} (color online) (a) Layer-resolved hole
count in an ideal structure (solid squares) and the hole count that
would obtain from structural roughness only (open circles). These
two distributions are convolved to achieve the hole distribution in
the real structure (panel c). The error bars represent our
uncertainty in determining $P_{\sigma}$ and $P_{\pi}$ in Eq. 1, as
well as statistical errors. Also shown is the nominal distribution
of Sr$^{2+}$ ions (open squares). (b) Sketch of the superlattice
hole distribution in an ideal structure, aligned to panel (a) for
comparison. (c) Sketch of the hole distribution in the real
structure, accounting for La/Sr interdiffusion. $p^{0}_{min}$ and
$p^{R}_{min}$ are the hole counts in the LCO layers contributed by
electronic effects and by roughness, respectively.}
\end{figure}

We wish to determine $p_{l}^{0}$, which is the distribution of
holes in the superlattice. This task is simplified by realizing that, by
symmetry (Fig.~\ref{fig:Figure3}b), our structure has only three inequivalent
$\rm CuO_{2}$ planes: the two central LSCO layers, the two outer LSCO layers,
and the two LCO layers. We denote the
hole occupancies in these layers by $p_{max}^{0}$, $p_{int}^{0}$ and
$p_{min}^{0}$, respectively (see labels in Fig.~\ref{fig:Figure3}b).
Further, by charge conservation, it must be that
$p_{max}^{0}+p_{min}^{0}+p_{int}^{0}=0.72$ holes. Therefore, to
completely determine the hole distribution in this superlattice, we
need only two more independent measurements of these three values.

The first measurement is the scattering near $L=2$. Written out, the
structure factor of this reflection is
\begin{align}
(\textbf{S}_2)_{ij} &= 0.84 \, \emph{x} \, [ f_{La}(\omega) -
f_{Sr}(\omega) ] \, \delta_{ij} +\nonumber
\end{align}
\begin{align}
+0.5( p_{min}^{0} + p_{max}^{0} - 2 p_{int}^{0}) f^D_{ij}(\omega)
\end{align}
\noindent where $x=0.36$ is the doping of the LSCO layers. The
origin $z=0$ was chosen between two LCO layers. This reflection is
not visible off-resonance, but becomes visible near the La $M_{5}$
edge because the difference $f_{La}(\omega)-f_{Sr}(\omega)$ becomes
large. However, near the MCP resonance, where $f^{D}_{ij}$ is large
(164 electrons at the resonance maximum \cite{ARSG2005}), the peak
is not observed. That it is absent is evidence that it is forbidden
by symmetry, i.e. $p_{max}^{0}+p_{min}^{0}-2p_{int}^{0}\approx 0$.
With the constraint of charge conservation this gives
$p_{int}^{0}\approx 0.24$. The nominal doping of this layer is 0.36
holes, so this result shows -- prior to knowledge of $p_{max}^{0}$
and $p_{min}^{0}$ -- that holes have diffused from the interface
layer.

The second measurement comes from the scattering near $L=1$. The
structure factor of this reflection is
\begin{align}
(\textbf{S}_1)_{ij} &= 1.66 \, \emph{x} \, [f_{Sr}(\omega) -
f_{La}(\omega) ] \, \delta_{ij} +\nonumber
\end{align}
\begin{align}
+ 0.87 \, (  p_{max}^{0} - p_{min}^{0}) f^D_{ij}(\omega).
\end{align}
\noindent Unlike the $L=2$ reflection, this peak is visible at both
the La $M_{5}$ and MCP resonances, indicating that it is allowed by
symmetry, i.e. the difference $p_{max}^{0}-p_{min}^{0}$ is nonzero.
Knowledge of this difference would completely determine the hole
distribution in the superlattice.

To determine $p_{max}^{0}-p_{min}^{0}$ directly from this
reflection, however would require a measurement of $\rm
\textbf{S}_{1}$ in absolute units, i.e. the knowledge of the overall
constant $A$ in Eq.~\ref{eqn:Equation1}, whose value derives from
many different effects. We can eliminate $A$, however, by noting
that $f^{D}_{ij}$ quickly goes to zero away from the resonance
energy~\cite{ARSG2005}, and that the remaining terms $f_{La}$ and
$f_{Sr}$ are tabulated~\cite{HGD1993}. $p_{max}^{0}-p_{min}^{0}$ can
therefore be determined from the relative increase in intensity of
$L=1$ scattering at MCP from that a few eV below the
edge.~\cite{Footnote1} A detailed energy dependence of $L=1$
scattering is shown in Fig.~\ref{fig:Figure2}. From the relative
increase in integrated intensity at MCP of $560~\%$, and using the
method outlined above to eliminate $A$, we obtain
$p_{max}^{0}-p_{min}^{0}=0.18$. From these constraints we calculate
$p_{max}^{0}=0.33 \pm 0.025$ and $p_{min}^{0}=0.15 \pm 0.025$.

This result is significant because it shows that the carriers are
not bound to the Sr$^{2+}$ ions, but rearrange between layers,
presumably to minimize their kinetic energy. For the perfect
structure (roughness effects will be considered below) the
``undoped" LCO layers have a hole count of at least 0.15 holes/Cu,
which is close to optimal doping for the LSCO system. This suggests
that superconductivity originates in the nominally insulating LCO
layers.

Having determined the hole distribution, it is useful to
characterize it with a screening length, $\lambda_{0}$. The simplest
way to define $\lambda_{0}$ is in linear Thomas-Fermi theory,
relating the charge $\rho^{ind}$ induced in a medium to the external
charge density $\rho^{ext}$ by
\begin{eqnarray}
\rho^{ind}(Q)=-\frac{k_{0}^2}{k_{0}^{2}+Q^{2}}\,\rho^{ext}(Q)
\label{eqn:Equation4},
\end{eqnarray}
\noindent where $k_{0}=1/\lambda_{0}$ is the Thomas-Fermi wave
vector.

In the present case the ``external" charge is that of the $\rm
Sr^{2+}$ ions~\cite{Footnote3}, and the induced charge is the hole
occupancy $p_{l}^{0}$, i.e.
$\rho^{ext}(Q)=\sum_{l}d^{Sr}_{l}\,e^{iQz_{l}}$ and
$\rho^{ind}(Q)=\sum_{l}p_{l}^{0}\,e^{iQz_{l}}$, where $d_{l}^{Sr}$
and $p_{l}^{0}$ were defined in
Eqs.~\ref{eqn:Equation1a}-\ref{eqn:Equation1b}. For the $L=1$
reflection, for which $Q=2\pi /c$, $\rho^{ext}=-1.66\, \emph{x}$ and
$\rho^{ind}=1.73\,(p^0_{max}-p^0_{min})$, resulting in $\lambda_{0}
=6.1\pm 2.0~\rm\AA$. This would be the characteristic size of the
accumulation region in a device made of these two materials.

We now consider the effect of interfacial roughness on the previous
conclusions.  The structure is not perfect, but contains step edges
from island formation during growth~\cite{G2008}, as well as some
La-Sr interdiffusion. For specular measurements such defects can be
modeled well by a convolution of the La-Sr profile with a Gaussian
roughness function. From convolution theorem this correction enters
as a multiplicative factor in momentum space
$R(Q)=e^{-Q^2\sigma^2/2}$, where $\sigma$ is the interface
roughness. This factor has the effect of suppressing higher order
reflections. From the ratio of the $L=1$ and $L=2$ reflections at
the La edge, we determine $\sigma=5.3\rm~\AA$. This roughness causes
a smearing of the hole density {\it in addition} to the electronic
redistribution we have already shown.  Including roughness, the
structure factor at $L=1$ is modified as
\begin{align}
(\textbf{S}_1)_{ij} = 1.66\,\emph{x}\, [f_{Sr}(\omega) -
f_{La}(\omega) ] \, \delta_{ij} \, R(2\pi/\emph{c}) +\nonumber
\end{align}
\begin{align}
+0.87 \,(p_{max}-p_{min}) f^D_{ij}(\omega),
\end{align}
where $p_{max}$ and $p_{min}$ are the actual hole occupancies in the
real, imperfect structure. From Eq.~\ref{eqn:Equation4} it follows
that $p_{max}-p_{min}=(p_{max}^{0}-p_{min}^{0})R(2\pi/\emph{c})$.
Thus in our earlier analysis, in which $S^D_1$ was determined by
normalizing to the value of $S_1^0$, the term $R(2\pi/\emph{c})$
divided out. As a result, the method used above to obtain
$\lambda_{0}$ is independent of roughness effects. Further, the
quantities $p^0_{max}=0.33$, $p^0_{int}=0.24$ and $p^0_{min}=0.15$
can be thought of as those that would have obtained if the structure
were defect-free. Including the roughness contribution, the true $p$
values in our structure are $p_{max}=0.30 \pm 0.03$, $p_{int}=0.24
\pm 0.03$ and $p_{min}=0.18 \pm 0.03$. We note that, if we include
only the roughness, the values are $p^R_{max}=0.35$,
$p^R_{int}=0.27$ and $p^R_{min}=0.10$. $p^{R}$ is defined as the
$\rm CuO_2$ layer doping due solely to the roughness of the
structure. Therefore, the dominant cause of doping of LCO layers is
electronic accumulation, not defects.

In conclusion, we have used RSXS to quantify the distribution of
holes in a superlattice of insulating LCO and nonsuperconducting
LSCO. We find that the distribution of holes differs from that of
the $\rm Sr^{2+}$ ions, indicating true charge accumulation. The
filling of the LCO layers is found to be close to 0.18 holes/Cu,
suggesting that the `insulating' layers are the main drivers of
superconductivity. Our study demonstrates that charge accumulation
can be achieved at transition metal oxide interfaces with existing
synthesis methods. Our study also demonstrates the usefulness of
RSXS for distinguishing atomic from electronic reconstruction at
transition metal oxide interfaces.

This work was supported by the Office of Basic Energy Sciences, U.S.
Department of Energy.  RSXS studies were supported by Grant No.
DE-FG02-06ER46285, with use of the NSLS supported under Contract No.
DE-AC02-98CH10886. Superlattice growth and characterization were
supported under Contract No. MA-509-MACA. Work in the FSMRL was
supported by grants DE-FG02-07ER46453 and DE-FG02-07ER46471.


\begin{thebibliography}{}

\bibitem{okamoto2004}
S. Okamoto and A. J. Millis, Nature \textbf{428}, 630 (2004).

\bibitem{okamoto2006}
S. Okamoto, A. J. Millis, and N. A. Spaldin, Phys. Rev. Lett.
\textbf{97}, 56802 (2006).

\bibitem{altieri}
S. Altieri, L. H. Tjeng, G. A. Sawatzky, Thin Solid Films
\textbf{400}, 9 (2001).

\bibitem{pickett}
R. Pentcheva and W. E. Pickett, Phys. Rev. Lett. \textbf{99}, 016802
(2007).

\bibitem{C2007}
J. Chakhalian \emph{et al.}, Science \textbf{318}, 1114 (2007).

\bibitem{OMGH2002}
A. Ohtomo \emph{et al.}, Nature \textbf{419}, 378 (2002).

\bibitem{RTCF2007}
N. Reyren \emph{et al.}, Science \textbf{317}, 1196 (2007).

\bibitem{ohtomo2004}
A. Ohtomo and H. Y. Hwang, Nature \textbf{427}, 423 (2004).

\bibitem{BMTW2008}
A. Bhattacharya \emph{et al.}, arXiv:0710.1452.

\bibitem{koida2002}
T. Koida \emph{et al.}, Phys. Rev. B \textbf{66}, 144418 (2002).

\bibitem{S2007}
S. Smadici \emph{et al.}, Phys. Rev. Lett. \textbf{99}, 196404
(2007).

\bibitem{BLVC2004}
I. Bozovic \emph{et al.}, Phys. Rev. Lett. \textbf{93}, 157002
(2004).

\bibitem{G2008}
A. Gozar \emph{et al.}, Nature \textbf{455}, 782 (2008).

\bibitem{Tsukazaki2007}
A. Tsukazaki \emph{et al.}, Science \textbf{315}, 1388 (2007).

\bibitem{note2}
The $L=3$ reflection is always absent by symmetry. This may readily
be verified by evaluating Eqs. 2-3.

\bibitem{ARSG2005}
P. Abbamonte \emph{et al.}, Nature Physics \textbf{1}, 155 (2005).

\bibitem{AVRS2002}
P. Abbamonte \emph{et al.}, Science \textbf{297}, 581 (2002).

\bibitem{Footnote5}
No change in the intensity of the $L=1$ reflection was seen across
the superconducting transition temperature.

\bibitem{Footnote4}
This quantity is equivalent to the ``raw" scattering factor defined
in Ref.~\cite{ARSG2005}.

\bibitem{A2006}
P. Abbamonte, Phys. Rev. B \textbf{74}, 195113 (2006).

\bibitem{HGD1993}
B.L. Henke, E. M. Gullikson, and J. C. Davis, At. Nucl. Data Tables
\textbf{54}, 181 (1993).

\bibitem{Footnote1}
The quantities $A$, $f_{La}$ and $f_{Sr}$ are all in principle
energy-dependent. However they do not change appreciably between
$\rm 520~eV$ and $\rm 530~eV$.

\bibitem{Footnote3}
Technically the sum includes all charged layers. However, only the
$\rm Sr^{2+}$ ions enter for $L=1$.

\end{thebibliography}
\end{document}